# Structural and magnetic properties of $Fe_{1+d}Te$ single crystals

Yoshikazu Mizuguchi[1*], Kentaro Hamada[a], Osuke Miura[a]

[a]*Tokyo Metropolitan University, 1-1 Minami-osawa, Hachioji, 192-0397, Japan*

**Abstract**

We have grown single crystals of $Fe_{1+d}Te$ by a conventional self-flux method. We obtained plate-like single crystals with $d_{nom} \geq 0.1$. The value of the magnetization increased with increasing excess Fe concentration, and a broadening of the antiferromagnetic transition was observed for $d_{nom} > 1.15$. Further, we noted that the antiferromagnetic transition of $Fe_{1.134}Te$ ($d_{nom} = 0.15$) was clearly suppressed to a lower temperature, which would indicate a possibility of controllability of magnetism by excess Fe concentration.

*Keywords*: iron telluride; Fe-based superconductivity; crystal growth; antiferromagnetism

**1. Introduction**

Since the discovery of superconductivity in the LaFeAsO system, tremendous efforts for elucidating the mechanism of Fe-based superconductivity, enhancing a transition temperature ($T_c$), and developing application as thin films and wires using Fe-based compounds have been actively performed [1-4]. Among the Fe-based superconductors, Fe chalcogenides, FeSe- and FeTe-based compounds, have the simplest crystal structure and composition [5-9], which should be great advantages for practical application. In fact, high-quality thin films of Fe chalcogenides have been successfully fabricated [10-16]. One of the most interesting properties of those thin films is a remarkable enhancement of $T_c$ by changing fabrication process and substrates. The $T_c$ of the $FeTe_{1-x}Se_x$ films clearly exceeds the value of the bulk materials when the crystal structure was optimized for appearing a high $T_c$ [14] as reported in pressure studies of Fe-chalcogenide superconductors [17-21].

Recently, we have reported fabrication of thin crystals of $FeTe_{1-x}S_x$ using the Scotch-tape method, which has been used in studies on graphene or another layered compound with a van-der-waals gap [22]. Thin crystals with a typical thickness of 40 nm were successfully fabricated by cleaving single crystals using Scotch tape. Further, we recently developed this technique, and realized fabrication of very thin $FeTe_{1-x}Se_x$ crystals with a typical thickness less than 2.5 nm corresponding to ~4 sheets of the $Fe_2(Te,Se)_2$ layer [23]. Thin crystals fabricated using this method will be suitable in both investigating basic physics of Fe-based superconductivity and fabrication of Fe-based superconducting devices.

Among the Fe chalcogenides, we are recently focusing on $Fe_{1+d}Te$, which is one of the parent compounds of the Fe-based superconductor and exhibits antiferromagnetic ordering below ~70 K. The antiferromagnetic ordering was suppressed and superconductivity was achieved by S- or Se-substitution at the Te site, or strain stress in $Fe_{1+d}Te$ thin films. Having considered these facts, we think $Fe_{1+d}Te$ thin crystal would have a potential to exhibit some interesting properties such as suppression of antiferromagnetic ordering and/or appearance of superconductivity. However, it is known that high-quality single crystals with a small Fe concentration, $d < 0.1$, cannot be obtained, while the high-quality crystals with large amount of excess Fe can be grown easily. To investigate physical properties of $Fe_{1+d}Te$ thin crystals and/or achieve FeTe-based superconducting devices, systematic studies on crystal growth, structural and physical properties of $Fe_{1+d}Te$ are required. Here we report crystal growth and characterization of structural and magnetic properties of $Fe_{1+d}Te$ single crystals which will be used in trials to fabricate $Fe_{1+d}Te$ thin crystals.

---

* Corresponding author. Tel.: +81-42-677-2748; fax: +81-42-677-2756.
  *E-mail address*: mizugu@tmu.ac.jp.

## 2. Experimental details

Single crystals of $Fe_{1+d}Te$ were prepared by a conventional self-flux method. Fe powders (99.9 %) and Te grains (99.999 %) with respective compositions, nominal $d$ ($d_{nom}$) = 0, 0.10, 0.15, 0.20, 0.25, 0.30, 0.40, were sealed into an evacuated quartz tube. The tube was sealed into an evacuated larger (outer) quartz tube to avoid reaction with the air due to crack in inner quartz tube during crystal growth. The double-sealed sample was heated at 1050 °C for 10 hours and cooled down to 650 °C with a cooling rate of -4 °C/hour. The obtained single crystals were observed using an optical microscope and a scanning electron microscope (SEM), and the actual composition of excess Fe ($d_{EDX}$) was investigated using energy dispersive x-ray spectroscopy (EDX). The crystal structure was characterized by x-ray diffraction (XRD) with a Cu-Kα radiation using a standard θ-2θ method. Temperature dependence of magnetization was measured using a superconducting quantum interference device (SQUID) magnetometer with an applied field of 1 T under after field-cooling (FC).

## 3. Results and discussion

Figure 1 shows the typical optical microscope images and SEM images. Although for $d_{nom}$ = 0 plate-like single crystals were not obtained, good plate-like crystals were obtained for $d_{nom}$ > 0.10. Among them, crystals with $d$ = 0.20 were the best quality, namely having clear surface and easily cleavable, as shown in the SEM image. We carried out EDX analysis to determine actual compositions. The composition was determined by averaging more than 5 data points. In Fig. 2, we plotted both the starting nominal composition and the actual composition. The minimum $d_{EDX}$ was ~0.09 and the maximum $d_{EDX}$ was ~0.2. At the intermediate region, $d_{EDX}$ changed with increasing nominal $d_{nom}$. This result is consistent with the previous investigations which suggested that the excess Fe concentration $d$ was $0.07 < d < 0.25$ [24-26]. On the basis of these results, we assumed that precipitates observed in the SEM image of the surface of $d_{nom}$ = 0 would be $FeTe_2$ and/or the other impurities, and that observed for $d_{nom}$ = 0.40 would be Fe particles.

Figure 3(a) shows the normalized XRD patterns for powdered crystals. Except for $d_{nom}$ = 0, containing impurity peaks of $FeTe_2$, all peaks were well indexed using the P4/nmm space group. The XRD patterns collected with single crystals exhibited only (00$l$) reflections as shown in Fig. 3(b). Lattice constants $a$ and $c$ were calculated using the peaks in Fig. 3(a), and plotted in Fig. 3(c, d) as a function of the $d_{EDX}$. The $c$ lattice constant seemed to decrease with increasing excess Fe concentration, while the $a$ lattice constant did not show a remarkable dependence on $d_{EDX}$. Finally we discuss the magnetic properties of $Fe_{1+d}Te$ single crystals. Figure 4(a) shows the temperature dependence of magnetization for $d_{nom}$ = 0.10, 0.15, 0.20, 0.30. The antiferromagnetic transition was observed around 70 K for $d_{nom}$ = 0.10. With increasing $d_{nom}$, the values of magnetization increased and the antiferromagnetic transition became broader as shown in Fig. 4(b). The increase of magnetization should be caused by the increase of excess Fe at the interlayer site, which has been reported to have magnetic moment, and/or the existence of the Fe impurities as observed on the surface of $d_{nom}$ > 0.2. The broadening of the transition should be related to the low-temperature crystal structure. For $Fe_{1+d}Te$ with small $d$, a structural transition from tetragonal into monoclinic occurs simultaneously with a commensurate antiferromagnetic transition. In contrast, with large $d$, the structural transition from tetragonal into orthorhombic occurs with an incommensurate antiferromagnetic transition [26]. This deference in the low-temperature structure would cause the broadening of the transition.

We observed an unexpected behavior for $d_{nom}$ = 0.15. It has been reported that the excess Fe concentration did not affect the transition temperature while that broadened the transition. However in our observation, the transition temperature clearly shifted to a lower temperature for $d_{nom}$ = 0.15. Further, interestingly, we can recognize two transitions: a broad transition around 60 K and a relatively sharp transition around 40 K. Almost the same data was reproduced using the other crystal with $d_{nom}$ = 0.15. Hence we are now confirming the details of this anomalous behavior. We think it is very interesting if the sharp (commensurate) transition was suppressed with increasing $d$ and the broad (incommensurate) transition was induced simultaneously. This assumption gives us with a possibility of suppression of the antiferromagnetism and appearance of superconductivity by only changing the excess Fe concentration and/or controlling the thickness of the thin crystals, for example by the Scotch-tape method.

## 4. Conclusion

We have grown single crystals of $Fe_{1+d}Te$ by a conventional self-flux method. We obtained plate-like single crystals with $d_{nom} \geq 0.1$. The actual composition was determined using EDX, and the controllable excess Fe concentration was found to be the same as the previous reports: $0.07 < d < 0.25$. The crystal structure was investigated by XRD, and a tendency that the $c$ lattice constant decreased with increasing excess Fe concentration was observed. Magnetization increased with increasing excess Fe concentration, and a broadening of the antiferromagnetic transition was observed for $d_{nom} >1.15$. Further, we noted that the antiferromagnetic transition of $Fe_{1.134}Te$ ($d_{nom} = 0.15$) was clearly suppressed to a lower temperature. This behavior would indicate the coexistence of the commensurate antiferromagnetic phase (Fe-poor phase) and the incommensurate antiferromagnetic phase (Fe-rich phase). Furthermore, the coexisting magnetic states might affect each other, and suppress the magnetic transition temperature to a lower temperature. To elucidate intrinsic nature of $Fe_{1+d}Te$, further investigations are required.

**Acknowledgements**

This work was partly supported by "Grant-in-Aid for Research Activity Start-up".

Figures

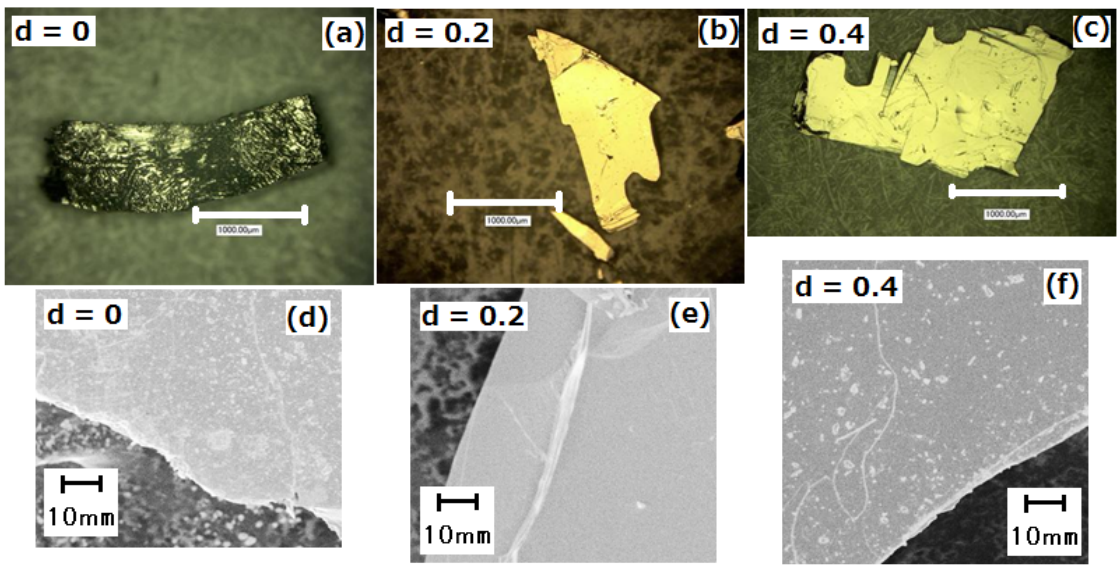

Fig. 1. (a-c) Optical microscope images and of $Fe_{1+d}Te$ crystals. (d-f) SEM images and of $Fe_{1+d}Te$ crystals.

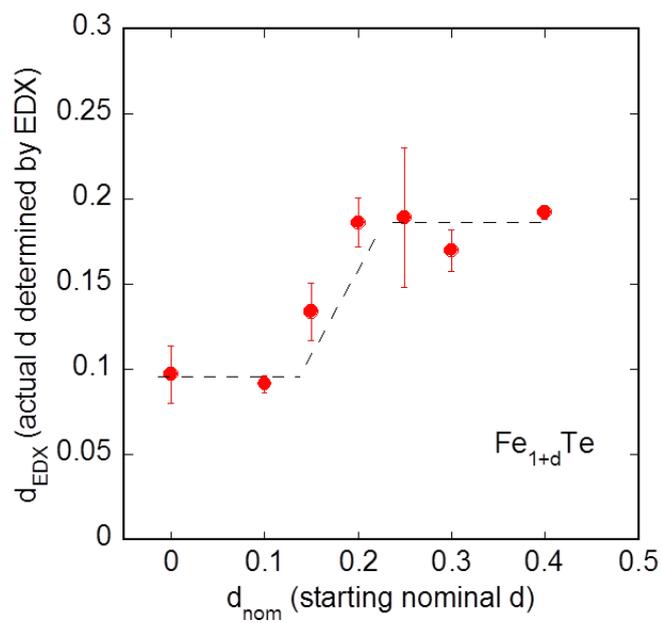

Fig. 2. $d_{nom}$ dependence of $d_{EDX}$.

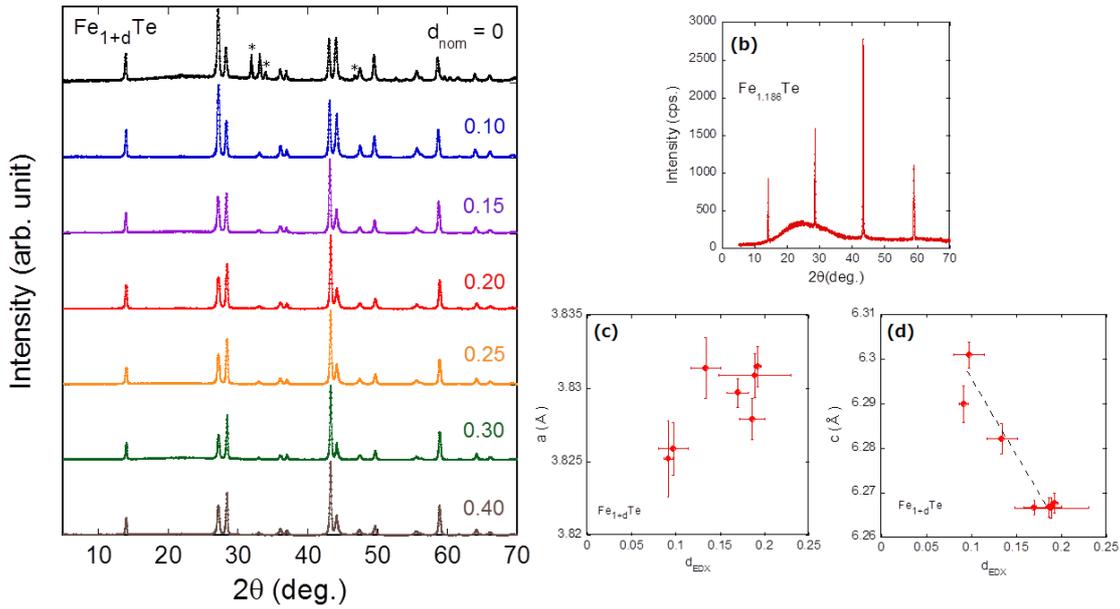

Fig. 3. (a) XRD patterns for the powdered $Fe_{1+d}Te$ crystals. Peaks of $FeTe_2$ impurity are indicated by asterisks. (b) XRD pattern for the $Fe_{1.186}Te$ single crystals. (c,d) Lattice constants $a$ and $c$ as a function of actual $d_{EDX}$.

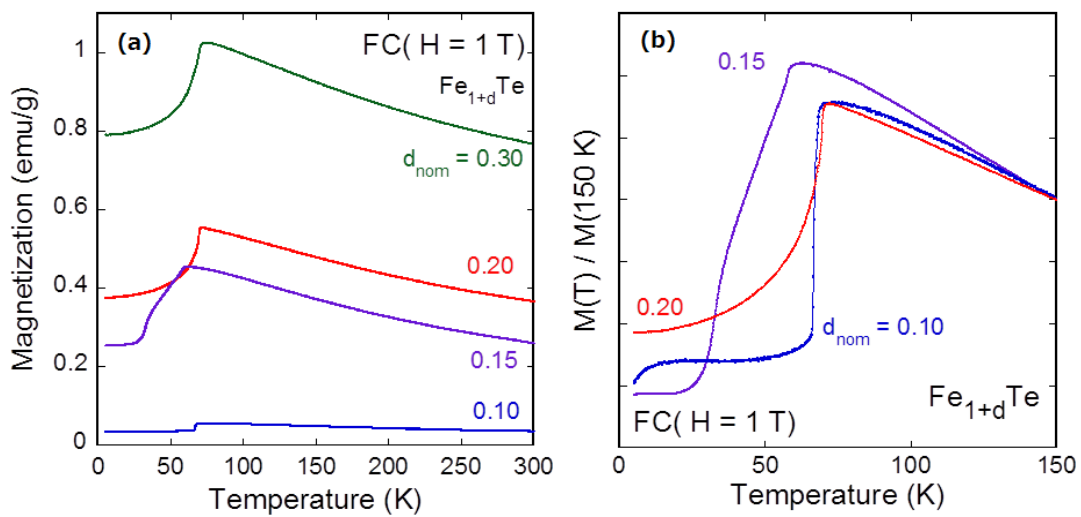

Fig. 4. (a) Temperature dependence of magnetization for $d_{nom}$ = 0.10, 0.15, 0.20 and 0.30. (b) Enlargement of the temperature dependence of magnetization normalized at 150 K.